# Photoemission study of PDI8-CN$_2$ /Au, PDI8-CN$_2$ / Si and PDI8-CN$_2$ - T6 interfaces

F.V. Di Girolamo[1], B. Mahns[2], M. Knupfer[2]

1 CNR-SPIN Naples, p.le Tecchio 80, 80125 Napoli, Italy

2 IFW- Dresden, P. O. Box 270116, D-01171 Dresden, Germany

We report on the study of the electronic properties in PDI8-CN$_2$ / Au, PDI8-CN$_2$ / Si, PDI8-CN$_2$ - T6 interfaces investigated by X-Ray (XPS) and Ultraviolet (UPS) Photoemission Spectroscopy. In order to verify the consistency with the results on field effect devices, heated substrates have been considered, while only in the case of PDI8-CN$_2$ / Au interface also sputter-cleaned Au has been used as a reference. We found a small interface dipole between PDI8-CN$_2$ and sputter-cleaned Au, which reverses its sign as a consequence of the "pillow effect" due to the presence of the contamination layer in the heated substrates. This result is in agreement with the transport measurements on PDI8-CN$_2$ field effect transistors, which show an "always on" behavior. We also found evidences of the formation of a built-in potential in T6 / PDI8-CN$_2$ and PDI8-CN$_2$ / T6 interfaces, mainly from the T6 side. This also well agrees with previous results on T6 / PDI8-CN$_2$ heterostructures devices, exhibiting transport properties explainable with the presence of an accumulation heterojunction at the interface.

I.  INTRODUCTION

The understanding of the physical phenomena occurring at the interfaces involving organic semiconductors has recently attracted considerable attention due to the different behavior exhibited with respect to their inorganic counterparts. It is sufficient to mention the small-molecule reorganization effect influence on the vacuum level alignment in metal / organic [1] and organic / organic interfaces [2], or the evidences of the formation of an accumulation heterojunction [3][4][5] rather than a depletion one in organic / organic heterostructures [6].

The determination of the electronic structure and the clarification of all the aspects regarding the energy level alignment in organic interfaces thus represents a necessary requirement for the understanding of the phenomena occurring in the heterostructures; to this aim, the most viable tool for a complete and quantitative investigation is Photoemission Spectroscopy [7] [8] [9].

In previous reports, transport measurements, Second Harmonic Generation (SHG) [10] and Ultraviolet Photoemission Spectroscopy (UPS) [11] suggested that the formation of a built-in electric field is able to explain the physical and electrical properties exhibited by sexithiophene (T6) / N,N'-bis (n octyl)-dicyanoperylenediimide ($PDI8-CN_2$) heterostructures, such as the increasing of the electrical performances of T6 transistors due to the doping effect of $PDI8-CN_2$.

The organic semiconductor, $PDI8-CN_2$, is a perylene diimide (PDIR) compound, which represents a class of n-type organic semiconductors with outstanding transport properties[12][13]. In fact, organic field effect transistors (OFETs) obtained using organic semiconductors belonging to PDIR family exhibit the highest n-type known mobilities [14] and high environmental stability [15]. Nevertheless, the understanding of the morphological and structural [16], transport [17] and electronic [18] properties of those materials is far from being complete.

In this study, the electronic properties of $PDI8-CN_2$ / Au, $PDI8-CN_2$ / Si, $PDI8-CN_2$ - T6 interfaces for both heated and sputter cleaned substrates (in the following $PDI8-CN_2$ / heated Au and $PDI8-CN_2$ / sputtered Au) have been investigated by X-Ray (XPS) and Ultraviolet (UPS) Photoemission Spectroscopy. It was shown that the effect of the contamination layer on the energy level alignment in PDI8-CN$_2$ / heated Au and PDI8-CN$_2$ / heated Si interfaces reverses the sign of the interface dipole, which can be an indication of a substrate workfunction change due to the contaminants, and which could make it easier to inject electrons in PDI8-CN$_2$ as also suggested by transport measurements [10] [12]. This effect does not occur for sputter cleaned Au. The shifts of the High Binding Energy Cutoff (HBEC), Highest Occupied Molecular Orbitals (HOMOs) and S2p XPS peak position also suggest the formation of an interface dipole and / or an electrostatic band bending at the interface between PDI8-CN$_2$ and T6, which is compatible with the hypothesis of the accumulation heterojunction [10] [11] at least in T6.

## II. EXPERIMENTAL

The x-ray photoemission spectroscopy (XPS) and valence-band ultraviolet photoemission spectroscopy (UPS) experiments were performed using an ultrahigh vacuum system (SPECS), which consists of three main parts: a chamber for measurement, a preparation chamber (both having base pressure of $1 \times 10^{-10}$ mbar) and a fast entry. The measurement chamber is equipped with a hemispherical electron energy analyzer PHOIBOS-150 (SPECS). A monochromatized Al Kα source (XR-50-M) provides photons with an energy of 1486.6 eV for XPS. Valence band measurements have been performed using a He discharge lamp (UVS-300) generating photons with an energy of 21.21 eV. The preparation chamber contains Knudsen-type evaporators and quartz microbalances.

The energy resolution of the photoemission experiments was determined by analyzing the width of the Fermi edge of a sputter-cleaned Au substrate to be 150 meV (UPS) and 350 meV (XPS).

Polycrystalline gold foils and n-type silicon wafers with a native oxide layer have been used as substrates for all the interfaces. In order to reproduce the experimental condition reported in Ref. [10] and Ref. [11] they have been cleaned through heating for 24 hours at 100°C.

As a reference, an argon-ion sputter-cleaned polycrystalline gold foil has also been used to grow PDI8-CN$_2$.

The cleanliness of the substrate surface and of the deposited films was checked using core-level x-ray photoemission spectroscopy. T6 (Sigma Aldrich) and PDI8-CN$_2$ (Polyera) films were grown via step-by-step evaporation of the respective materials onto the substrates kept at 100°C. Deposition rates calibrated with the quartz microbalance were set to 0.2 and 0.1 nm/min for PDI8-CN$_2$ and T6, respectively. After each T6 or PDI8-CN$_2$ deposition step, C 1$s$, S 2p, N 1s, O 1s, Au 4$f$ (only for gold substrates) and S 2p (only for silicon substrates) core – level and valence band photoemission spectra were acquired in the normal emission geometry in order to follow changes of the electronic structure upon formation of the interfaces.

The energy scale for the XPS measurements was calibrated to reproduce the binding energy (BE) of the Au 4$f_{7/2}$ core level (84.0 eV). The UPS data were corrected accounting for contributions of He I$_\beta$ and He I$_\gamma$ satellites. It was assumed that these contributions amount to 1.8% (He I$_\beta$) and 0.5% (He I$_\gamma$) of the He I$_\alpha$ intensity and have the same line shapes as the He I$_\alpha$ signal. The He I$_\beta$- and He I$_\gamma$- derived subspectra were shifted relative to the He I$_\alpha$ spectrum toward lower BE's by 1.87 and 2.52 eV, respectively, and subtracted. The spectral line shapes related to the bare T6 and PDI8-CN$_2$ photoemission signals remain virtually unchanged at all stages of deposition.

All UPS measurements have been done by applying a sample bias of -5 V to obtain the correct, sample determined, HBEC. For further details regarding determination of interface parameters using Photoemission see for instance [7], [19].

## III. PHOTOEMISSION STUDY OF HEATED Au /PDI8-CN$_2$, HEATED Si /PDI8-CN$_2$, SPUTTER-CLEANED Au / PDI8-CN$_2$ INTERFACES

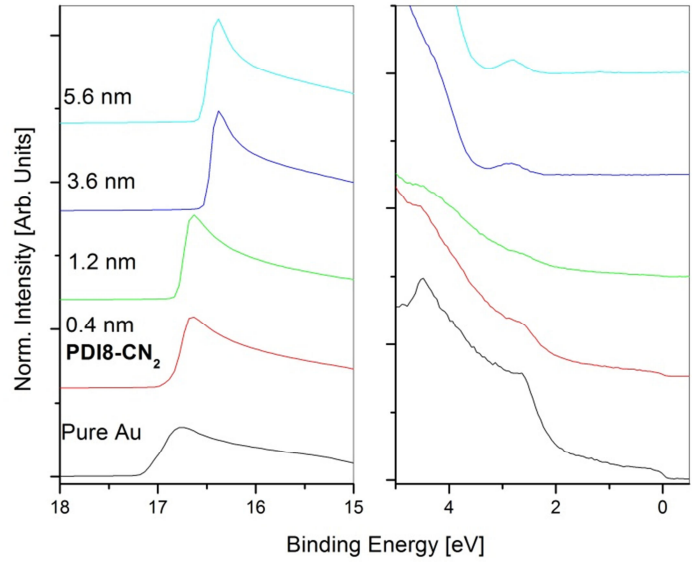

a)

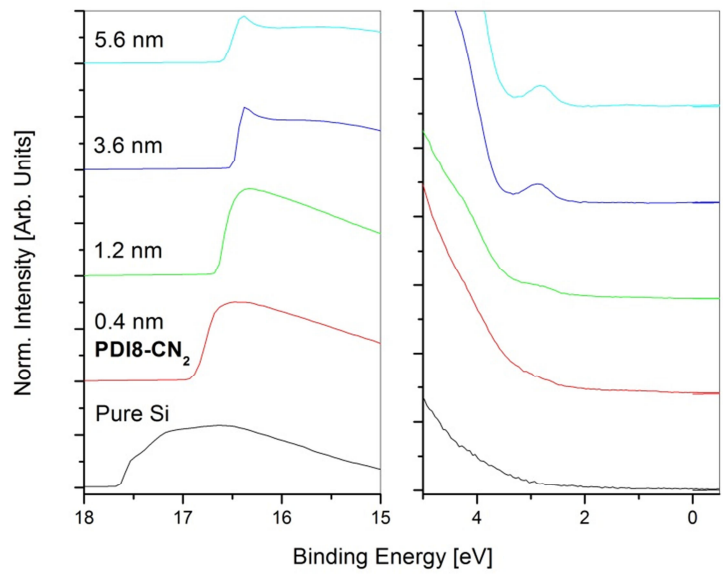

b)

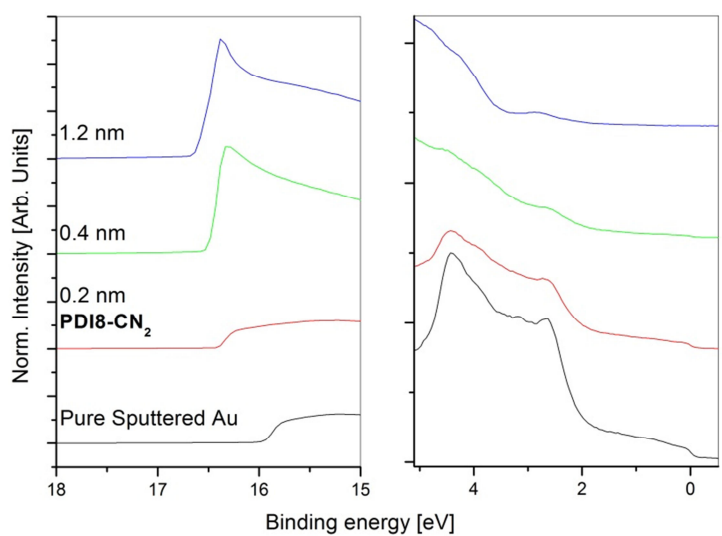

c)

Fig. 1. UPS spectra (He Iα) of PDI8-CN$_2$ for increasing thicknesses deposited on a) heated Au, b) heated Si, c) sputter-cleaned Au. UPS spectra of the pure substrates are also shown to evidence the Fermi level position.

In figure 1 the UPS spectra of PDI8-CN$_2$ / heated Au ( a ), PDI8-CN$_2$ / heated Si (b), PDI8-CN$_2$ / sputter-cleaned Au (c) interfaces are reported for increasing thickness of the organic layer. In all the cases, the substrate features disappear for increasing thicknesses, while PDI8-CN$_2$ ones become evident. In particular, the HOMO peak is already visible at 1.2 nm for sputter-cleaned gold as a substrate, while it appears only at 3.6 nm for heated substrates. The position of the HOMO level (around 2.1 eV) does not show a significant shift with thickness. On the other hand, a shift of the HBEC is evident in all cases virtually without any binding-energy change, indicating the formation of an interface dipole; nevertheless, while the behavior of PDI8-CN$_2$ / heated Au and PDI8-CN$_2$ / heated Si interfaces is comparable, the PDI8-CN$_2$ / sputter-cleaned Au interface represents a different case.

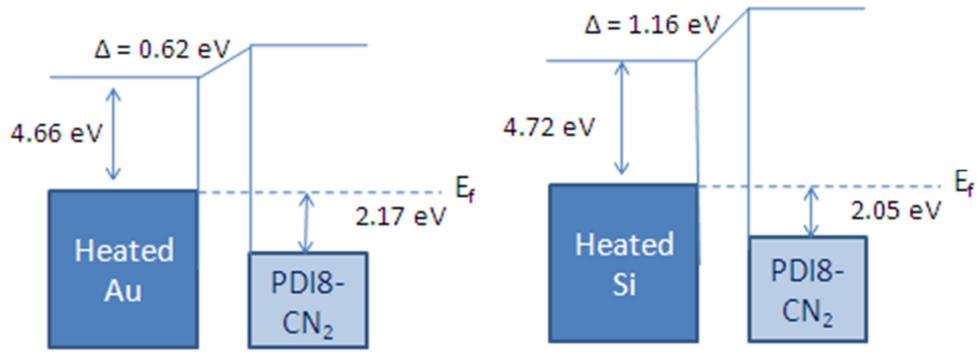

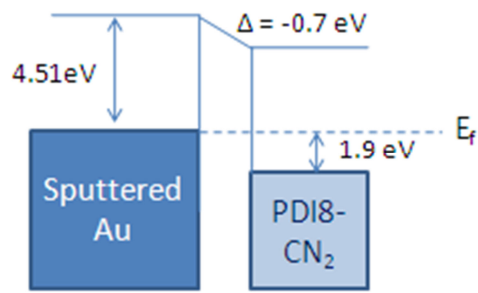

c)

Fig. 2. Schematic energy-level diagram of a) PDI8-CN$_2$/ heated Au, b) PDI8-CN$_2$/ heated Si, c) PDI8-CN$_2$ / sputter-cleaned Au interfaces as obtained from photoemission spectroscopy investigations.

The PDI8-CN$_2$ / heated Au, PDI8-CN$_2$/ heated Si, PDI8-CN$_2$ / sputter-cleaned Au interface dipoles were consequently extracted subtracting the substrate HBEC from the organic HBEC (as reported in the literature [7] [20]) and reported, respectively, in the Figure 2 a), b) and c). In particular, as summarized in the figure 1, the High Binding Energy Cutoff (HBEC) decreases at increasing PDI8-CN$_2$ thickness for heated substrates (in agreement with previous results[11]), while it increases for sputter-cleaned Au. The origin of this occurrence can be addressed to a strong reduction of the interface dipole due to the presence of a "pillow layer" of contaminants between the organic and the heated substrates [7] [20], as will be discussed below. Our XPS core-level analysis indeed reveals that the heated Au surfaces are covered by a contamination layer that consists of carbon, oxygen and nitrogen. Heated Si surfaces were instead only contaminated by carbon and oxygen. As reported in the literature [7] [20] the contaminant layer causes a decrease in the interface dipole, which in our case even reverses the sign of the dipole. A similar effect was already reported for PEDOT-PSS / pentacene and PEDOT-PSS / parasexiphenyl interfaces [1] and can be attributed to the so-called "pillow effect" [1]: the Coulomb repulsion of the electronic density of the organic molecule and the surface metal electrons locally suppresses the tail of electron wave function that spills into vacuum, reducing the work function of the metal. Work function changes due to adsorbates on metal surfaces have been extensively investigated, in particular for noble gases on series of metals [21].

IV. PHOTOEMISSION STUDY OF T6 - PDI8-CN$_2$ INTERFACES ON HEATED Au AND Si

IV.I. XPS MEASUREMENTS

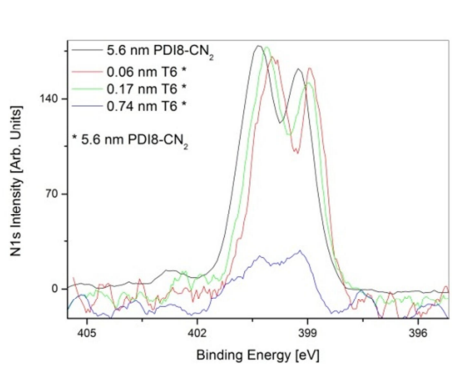

a)

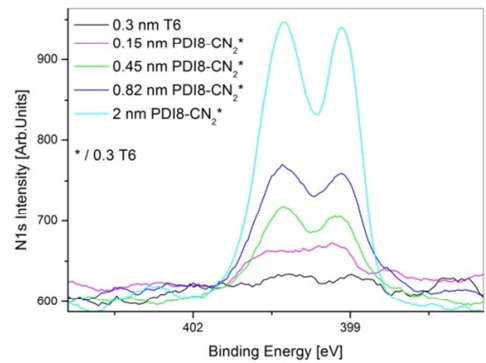

b)

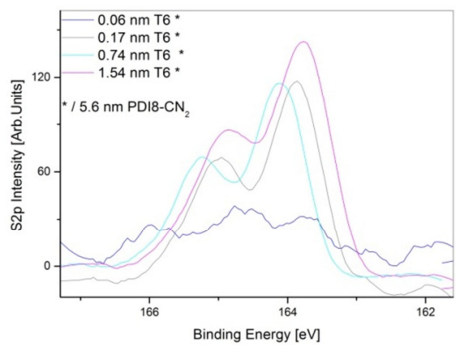

c)

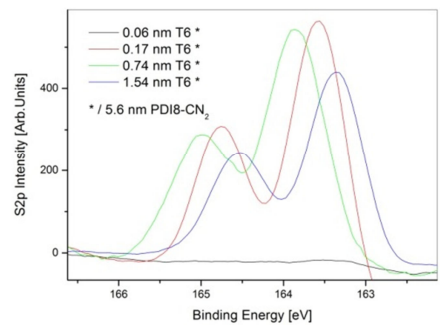

d)

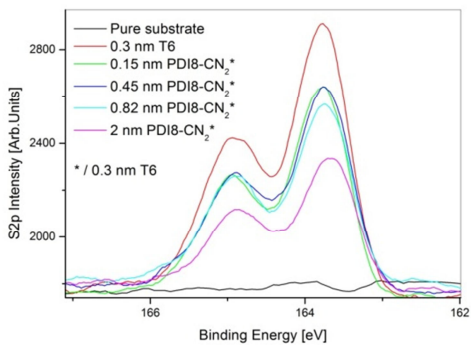 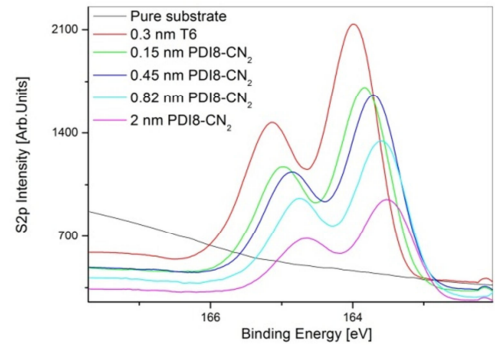

e)                                                                                                          f)

Fig. 3. N1s ( a , b ) and S2p ( c, d, e, f ) XPS spectra obtained upon deposition of T6 on PDI8-$CN_2$ ( a, c, d ) and upon deposition of PDI8-$CN_2$ on T6 ( b, e, f ) respectively on heated Si (a,b,d,f) and heated Au (c, e).

In Figure 3 the evolution of the N1s (a, b) and S2p (c, d, e, f) core level peaks are reported, respectively indicative of the presence of PDI8-$CN_2$, and of T6. Since nitrogen is also present in the Au contamination level, the N1s peak evolution is shown for clarity only for silicon substrates.

For the T6 / PDI8-$CN_2$ interface (Figure 3 a) the deposition of T6 clearly suppresses the N1s peak intensity, as a consequence of the decrease of the nitrogen content at the surface. Conversely, in PDI8-$CN_2$ / T6 interfaces (Figure 3 b) the N1s peak intensity increases as the PDI8-$CN_2$ thickness increases, as expected.

In an analogous way, the evolution of the S2p core level peak intensity directly follows the T6 deposition: for T6 / PDI8-$CN_2$ it increases at increasing T6 thicknesses, both on heated Au (Figure 3 c) and heated Si (Figure 3 d); for PDI8-$CN_2$ / T6 interfaces it decreases at increasing T6 thicknesses, also in this case both on heated Au (Figure 3 e) and heated Si (Figure 3 f).

In Figure 3 f) a monotonous shift of the S2p peak position with PDI8-$CN_2$ thickness is also evident, which is an indication of a binding energy shift and suggests a band bending phenomenon in T6. This finding will be further discussed in the following section.

## IV.II. UPS MEASUREMENTS

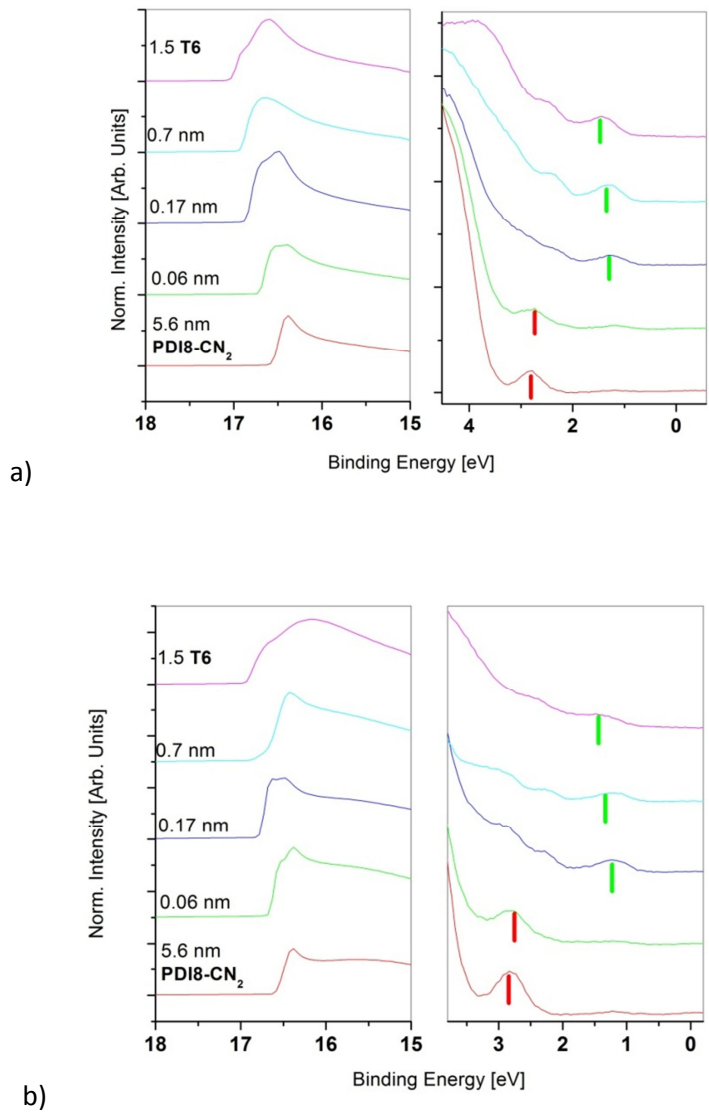

a)

b)

Fig. 4. UPS spectra (He Iα) of T6 / PDI8-CN$_2$ heterostructure for increasing T6 thicknesses deposited on a) Au and b) Si. UPS spectra of the sole PDI8-CN$_2$ is shown for comparison. The PDI8-CN$_2$ HOMO peak position is evidenced by a red vertical line, the T6 HOMO by a green vertical line.

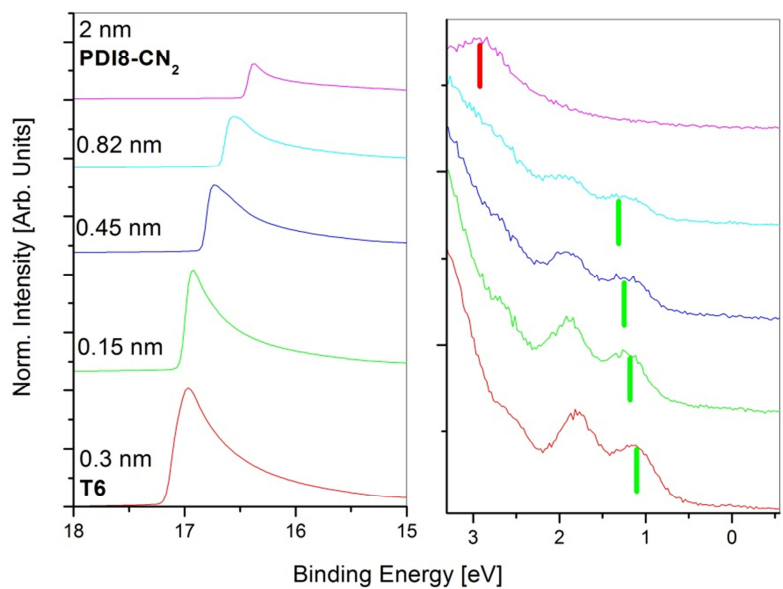

a)

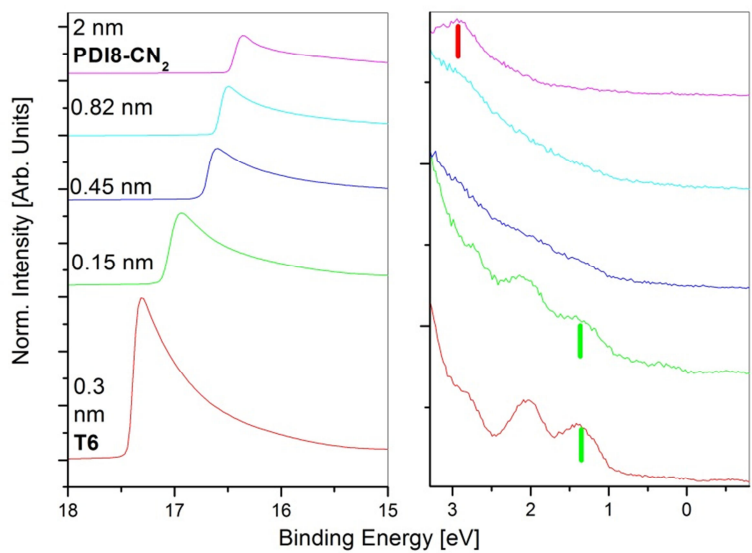

b)

Fig. 5. UPS spectra (He Iα) of PDI8-CN$_2$ / T6 heterostructure for increasing PDI8-CN$_2$ thicknesses deposited on a) Au and b) Si. UPS spectra of the sole T6 is shown for comparison. The PDI8-CN$_2$ HOMO peak position is evidenced by a red vertical line, the T6 HOMO by a green vertical line.

In Figure 4 and Figure 5 the UPS measurements respectively for the T6 / PDI8-CN$_2$ and the PDI8-CN$_2$ / T6 interfaces are reported. The PDI8-CN$_2$ HOMO peak position is evidenced by a red vertical line, the T6 HOMO by a green vertical line.

In Figure 4 T6 features slowly become more and more evident with increasing T6 thickness, as expected. The final position of the T6 HOMO peak is consistent with the bulk value and, in general, the T6 electronic features are in agreement with the ones reported in the literature [11][20]. A shift of the HBEC towards higher binding energies also evidences the change in the vacuum level at the T6 / PDI8-CN$_2$ heterojunction due to the alignment of the energy levels consequent to the formation of the interface. These results are irrespective of the substrate, indicating that all the further considerations can be considered to be only due to an interfacial (T6 - PDI8-CN$_2$) effect.

Analogously, in Figure 5 T6 features disappear at increasing PDI8-CN$_2$ thickness; PDI8-CN$_2$ features become visible only at high coverages; the HOMO position reaches a value which is close to the bulk one. The shift of the HBEC, this time towards lower binding energies, indicates the energy level alignment at the PDI8-CN$_2$ / T6 interface; the opposite direction of the shift suggests that the built-in potential at the interface has an opposite sign respect to the case of the reverse configuration. In this case we can observe that the results are not completely irrespective of the substrate: even if the position of the peaks is basically the same for the pure T6, the 45 s PDI8-CN$_2$ on T6 and the pure PDI8-CN$_2$ on heated Au and heated Si, the absence of peaks in the PES of the other two interfaces makes the same comparison impossible. It was already reported that a low intensity of the HOMO peak can be attributed to an island growth [11], which is typical of T6 [22]. Moreover, this indicates that the electronic properties of T6 depends on the substrate and on the specific molecular packing [23], while for PDI8-CN$_2$ they depend only on the presence of a contaminant layer but not on the substrate.

The shift of both the HOMO peak position and of the HBEC suggests that not only a dipole but also a band bending phenomenon occurs at the interface. Moreover, the overall picture is consistent with the hypothesis of the formation of an accumulation heterojunction at the interface, as also suggested in previous reports [10] [11].

IV.III. ANALYSIS OF THE XPS AND VALENCE BAND SPECTRA

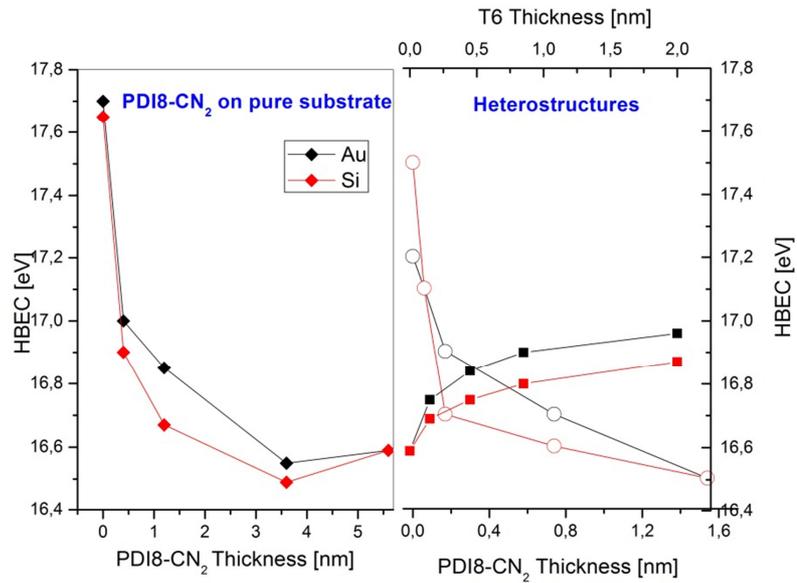

Fig. 6. HBEC onset positions in the PES spectra upon formation of PDI8-CN$_2$ / Au, PDI8-CN$_2$ / Si, PDI8-CN$_2$ / T6 and T6 / PDI8-CN$_2$ interfaces.

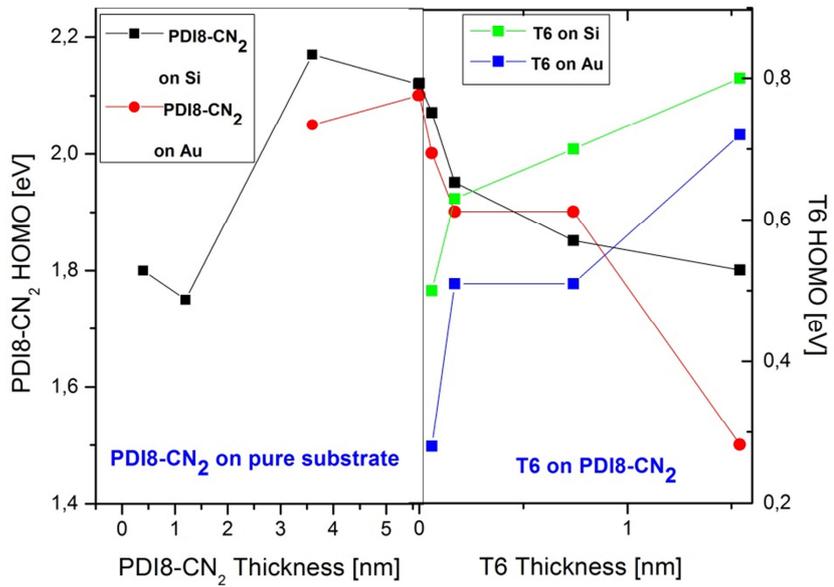

Fig. 7. HOMO onset positions in the PES spectra upon formation of PDI8-CN$_2$ / Au, PDI8-CN$_2$ / Si and T6 / PDI8-CN$_2$ interfaces.

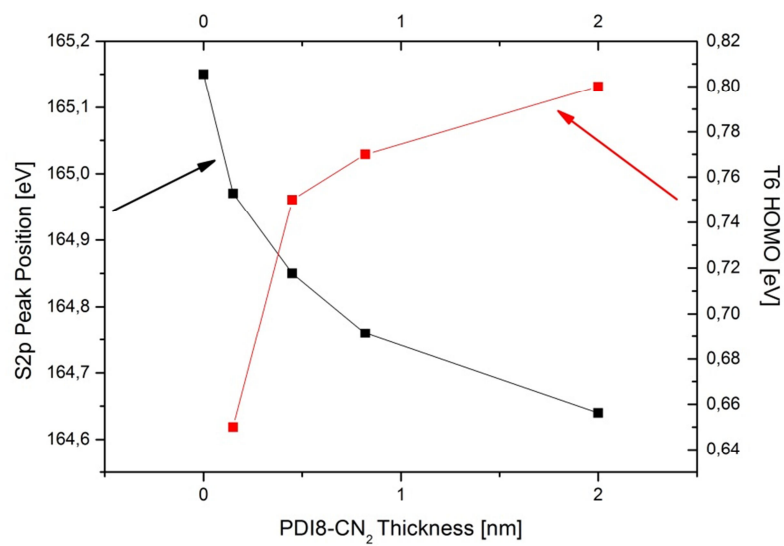

Fig. 8. S2p peak position upon formation of T6 / PDI8-CN$_2$ interface compared with the HOMO onset positions in the PES spectra upon formation of PDI8-CN$_2$ / T6 interface.

In this section the HBEC (Figure 6) and HOMO (Figure 7) onset positions are extracted from the PES spectra upon formation of PDI8-CN$_2$ / Au, PDI8-CN$_2$ / Si, PDI8-CN$_2$ / T6 and T6 / PDI8-CN$_2$ interfaces on both heated Au and heated Si, in order to evidence the changes in the interfacial electronic structure. For the extraction of the HOMO onset, the peak position evidenced by a red and a green vertical line in the Figure 4 were respectively used for PDI8-CN$_2$ and T6.

It is necessary to mention that the extraction of the HOMO onset for the PDI8-CN$_2$ / T6 interface was not always possible (because of the absence of the HOMO peak for some interfaces, as discussed in the previous section) and consequently the values are not reported in Figure 7. Nevertheless, T6 HOMO onset values extracted from the peaks evidenced by a green vertical line in Figure 5 are used in Figure 8 as a comparison with the S2p peak position extracted from the XPS measurements for the same (n.b. PDI8-CN$_2$ / T6)interface.

Indeed, as discussed in the previous section, in the XPS measurements on the T6 / PDI8-CN$_2$ interface a monotonous shift in the S2p peak position is evident on heated Si. In principle those changes are also reflected in an energy shift of corresponding core levels and consequently support the hypothesis of a band bending phenomenon at the interface.

The evolution of the HBEC onset position in the PES spectra evidences the formation of an interface dipole and/or a built-in potential firstly at the interface between PDI8-CN$_2$ and the pure substrate and then at the T6 / PDI8-CN$_2$ heterostructures. It is interesting to evidence that the decrease of the HBEC at increasing PDI8-CN$_2$ thickness and its increase upon deposition of T6 on PDI8-CN$_2$, are in agreement with previous reports [11] and, in the latter case, with the hypothesis of the formation of an accumulation junction. It is worth to mention that also the decrease of the HBEC at increasing PDI8-CN$_2$ thickness in the PDI8-CN$_2$ / T6 interface also supports the same conclusion.

Regarding the evolution of the HOMO peak, we can observe that, on pure substrates, it almost immediately pins at the saturation value (around 2.1 eV), corresponding to the bulk one. This, along with the observation of the shift of the HBEC, suggests that the built-in field at the interface is mainly due to the changes in the PDI8-CN$_2$ layer structure during the deposition, rather than a band bending phenomenon (which is indeed not evidenced by a shift in the HOMO as soon as the bulk structure is formed).

When the T6 / PDI8-CN$_2$ interface is formed, the shift in the HBEC is accompanied by a monotonous change in both the HOMO positions (i.e. upward shift of the T6 HOMO while downward shift of the PDI8-CN$_2$ one), which suggests that in this case also a band bending phenomenon occurs.

On the other hand, when considering PDI8-CN$_2$ / T6 interface, the results are less clear due to a not pronounced or even not present HOMO onset position. In this case, indeed, is not possible to verify if the PDI8-CN$_2$ HOMO undergoes a shift, even if T6 HOMO seems to shift upwards as reported for T6 / PDI8-CN$_2$ interface (at least on heated Au substrate), see Figure 8. By the way, in this case, a further evidence of the band bending occurring in T6 is given by the shift of the S2p peak position on heated Si, as reported in Figure 8.

The experimental results suggest that the electronic structure is not irrespective of the deposition order. In the case of T6 as a lower layer, it seems to be dependent on the substrate, as shown from the comparison of Figure Figure XPS e) and XPS f) and the comparison of 5 a) and 5 b), contrarily to PDI8-CN$_2$.

These observations can be explained by considering the different morphology exhibited by T6 and PDI8-CN$_2$: the first one indeed shows mostly an island growth for the early stages of deposition which evolves into a layer by layer one at higher thicknesses [22] while the second one is characterized by flat layers both in the bulk and in thin films [Liscio]. Moreover, photoemission studies have clearly demonstrated a strong variation of the T6 electronic structure with the molecular packing [23]. These observation suggest that T6 electronic properties are strongly influenced by morphology, while this seems to be less true for PDI8-CN$_2$ or, conversely, that T6 morphology is more influenced by the growth conditions and the substrate respect to PDI8-CN$_2$. Since in heterostructures different growth conditions and substrates are necessarily used, this explains why the results are not irrespective of the deposition order.

## V. CONCLUSION

In the present paper, a study of the electronic properties of heated PDI8-CN$_2$ / heated Au, PDI8-CN$_2$ / heated Si, PDI8-CN$_2$ / sputter-cleaned Au, T6 / PDI8-CN$_2$ and PDI8-CN$_2$ / T6 interfaces by X-Ray (XPS) and Ultraviolet (UPS) Photoemission Spectroscopy is presented.

The results evidence that PDI8-CN$_2$ electronic properties are strongly affected by the presence of a contamination layer at the interface with the substrate more than by the substrate itself. Since the contamination layer causes a strong decrease of the interface dipole (even reversing its sign) but does not influence the molecular packing, this suggests that PDI8-CN$_2$ electronic properties are not significantly affected by the molecular packing itself.

Regarding the organic / organic interfaces, it is shown that while the behavior for T6/PDI8-CN$_2$ interface (where the T6 surface morphology evolves on PDI8-CN$_2$) is the same irrespective of the substrate, the PDI8-CN$_2$ / T6 interface (where the T6 morphological changes on different substrates counts) depends on the substrate. This suggests that the properties at the interface are strongly dependent on the T6 morphology.

In any case, some evidences are given supporting the hypothesis of a band bending at the T6-PDI8-CN$_2$ interface, at least from the T6 side, which agrees with the hypothesis of an accumulation heterojunction.

Acknowledgments

We thank M. Naumann, A.Schubert, R. Hübel and S. Leger for technical assistance. This work has been supported by the Deutsche Forschungsgemeinschaft (grant number KN393/14) and by Program FP7-RegPot2010-1FP7, MAMA project.